\documentclass[letterpaper,14pt]{article}
\usepackage[centertags]{amsmath}
\usepackage{amsfonts}
\usepackage{amssymb}
\usepackage[english]{babel}
\usepackage[dvips]{graphicx}
\newcommand{\ud}{\mathrm{d}}

\begin{document}\begin{center}
{Zero-point length, extra-dimensions and string $T$-duality}
\end{center}
\begin{center}
{Euro Spallucci and Michele Fontanini}
\end{center}

\begin{center}Department of Theoretical Physics,
         University of Trieste, Strada Costiera 11, 34014 Trieste,
         Italy, and Sezione INFN di Trieste,
         Strada Costiera 11, 34014 Trieste, Italy
\end{center}
\vskip 1cm

\begin{center}
\begin{abstract}
A fully satisfactory Unified Theory of all fundamental interactions, including
gravity, is still to come. The most promising candidate to this role is Super
String Theory, which has totally changed our perspective about space, time and
matter since its \textit{revolutionary} re-appearance in the 80's.\\ 
So far, String Theory represents  the \textit{only} technical and conceptual 
framework, available, allowing consistent perturbative quantum gravity 
calculations, and selecting $SO(32)$ and $E_8\otimes E_8$ as anomaly-free 
grand unification gauge groups.\\
However, many non perturbative problems, still remain unsolved.
Anyway, whatever String Theory really means in a non-perturbative sense,
some fundamental spacetime features already emerge in a  clear and well 
definite way:
\begin{itemize} 
\item spacetime must be \textit{higher dimensional}, with
(~geometrically/topolo\-gically~) non-trivial space-like (~large/small~)
extra-dimensions;
\item there is a ``\textit{minimal length}'' below which the very concept
of spacetime continuum becomes physically meaningless.
\end{itemize} 
The existence of a minimal length, or ``\textit{zero-point
length}'', in  four-dimen\-sio\-nal spacetime has 
been often conjectured as a necessary, and welcome, aftermath of any
correct  quantum  theory of gravity. The presence of such a length should act 
as a \textit{universal} $UV$ regulator in any low energy quantum field theory 
solving the longstanding problem of $UV$ divergences.\\
In this paper, we are going to put in a single consistent framework apparently
unrelated pieces of information, i.e. zero-point length, extra-dimensions,  
string $T$-duality.
More in details we are going to introduce a modified Kaluza-Klein theory
interpolating between (~high-energy~) string theory and (~low-energy~) quantum 
field theory. In our model zero-point length is a four dimensional 
``virtual memory'' of compact extra-dimensions length scale. Such a scale 
turns out to be determined by $T$-duality inherited from the underlying 
fundamental string theory. 
From a low energy perspective short distance infinities are cut off by
a minimal length which is proportional to the square root of the string
slope, i.e. $\sqrt{\alpha^\prime}$. Thus, we bridge the gap between the
ultra-relativistic string domain and the low energy arena of point-particle
quantum field theory.
\end{abstract}
\end{center}

\section{Introduction}

Our understanding of the basic principle of physics is encoded into three 
fundamental constants: the speed of light $c$, the Planck constant $\hbar$,
the Newton constant $G_N$. Each constant labels a different class of
phenomena described by a proper physical theory:\\
$c$ $\rightarrow$ Special Relativity;\\
$\hbar$  $\rightarrow$ Quantum mechanics;\\
$G_N$ $\rightarrow$ Gravity.\\
By coupling in various ways these constants we get the cornerstones of modern 
physics:\\
i) $c$ $+$ $\hbar$ $\rightarrow$ Relativistic Quantum Mechanics, or 
Quantum Field Theory;\\
ii) $c$ $+$ $G_N$ $\rightarrow$ General Relativity. \\
Finally, by combining together i)+ ii)\footnote{ The combination
$\hbar$ $+$ $G_N$ is not physically meaningful as it would correspond to
a non-relativistic quantum theory of gravity in disagreement with
the relativistic character of General Relativity. } 
we get:\\
$c$ $+$ $G_N$ $+$ $\hbar$ $\rightarrow$ Quantum Gravity, or String-Theory.\\
Is there any other fundamental constant missing in the scheme above?\\
Even if a general consensus is missing, many clues point out that
the correct ``fundamental triplet'' is  $c$, $\hbar$, and $l_0$ in
place of $G_N$ \cite{yoneya}, where $l_0$ is
a \textit{fundamental}, or  \textit{minimal} length characterizing the very 
short distance spacetime geometry. In other words, $l_0$ is the
``\textit{quantum}'' of length. Replacing $G_N$ with $l_0$ is
much less trivial than it could appear at a first glance, as the
presence of a fundamental length must not conflict with Special Relativity.
That means $l_0$ must be \textit{inert} under Lorentz transformations in order 
to represent the same fundamental quantum of length for any inertial observer.
Early attempts to build up a quantized spacetime date back to Snyder
work in `47 \cite{snyder}  and were motivated by the need to modify quantum
fields point interactions in order to avoid ultraviolet divergent amplitudes. 
However, the overwhelming success of renormalization theory and the impressive
predictions of QED put aside Snyder approach. His efforts were largely ignored
up to the recent revival of interest for non-commutative geometry. Indeed,
the most straightforward way to quantize spacetime is promoting coordinates
to the role of operators and require an appropriate commutation relation:
\begin{equation} 
x^\mu\longrightarrow \mathbf{x}^\mu\quad :\quad \left[\, \mathbf{x}^\mu\ ,
\,\mathbf{x}^\mu\,\right]= i\, \theta^{\mu\nu}
\end{equation} 
It follows from dimensional consideration that $\theta^{\mu\nu}$ has dimensions
of a $(\, \mathit{length}\, )^2$. Thus,
\begin{equation} 
\theta^{\mu\nu}=l_0^2\, \vartheta^{\mu\nu}
\end{equation}
However, if $\theta^{\mu\nu}$ transforms as a rank-two antisymmetric 
tensor under Lorentz transformations
is not clear at all why the length scale $l_0$ should  remain invariant. In
recent times, we showed that only if all the components of $\theta^{\mu\nu}$ 
are equal in the block-diagonal basis it is possible to define $l_0$ in a 
Lorentz invariant way \cite{ncqft}.
Originally, Snyder solved the problem in a different way: he started from a
five dimensional Lorentz algebra and identified non-commuting four-dimensional
coordinates with $M^{5\,\mu}$ algebra generators. We are not going into 
the details of Snyder approach, rather we would like to remark two features:\\
i) the motivation was to cure ultraviolet divergences in quantum field
theory;\\
ii) compatibility between a minimal length and Special Relativity requires
an higher dimensional spacetime. \\
Let us elaborate a bit issue i). 
Conventional quantum field theory exhibits ultraviolet (~UV~) divergences.  
It was always  thought that this unphysical behavior is a consequence
of the conflict between quantum matter  and classical spacetime geometry.
A fully quantized theory of space, time and matter is expected to be able
to ``cure'' the problem \cite{garay}. 
This is to 
be expected in that one cannot operationally define
scales below Planck scale in any model that incorporates the principles of 
quantum theory and gravity \cite{padma1}, \cite{balbi}.
This makes Planck length $L_{Planck}\approx 10^{-33}\, cm$
acting as the ``zero-point length'' of the spacetime. 
The propagator in a field theory should therefore 
be ultraviolet finite when quantum gravitational corrections are incorporated.
Such a conclusion follows from two general features of quantum field theory
interacting with gravity:\\
i)gravity is a non-polynomial interaction with a universal coupling
to any kind of matter field;\\
ii) non-polynomial interactions produce a damping effect on the large momentum 
behavior of the propagator.

To clarify the relation between non-polynomial interactions and the presence
of a fundamental length-scale \cite{fradefi}, we will consider a massless scalar field
and suppose that some interaction is able
to modify the short distance behavior of the Green function as follows:
\begin{equation}
G_0\left(\, x-y\,\right)=-\frac{1}{4\pi^2}
\frac{1}{ \left(\, x-y\,\right)^2 }\longrightarrow 
G\left(\, x-y\ ; l_0\,\right)=-\frac{1}{4\pi^2}
\frac{1}{\left(\, x-y\,\right)^2 + l_0^2}
\label{greenl0}
\end{equation}
where $l_0$ represents a quantum mechanical non-vanishing \textit{residual} 
length in the coincidence limit $x\to y$. The presence of $l_0$ in
(\ref{greenl0}) corresponds to the introduction of a \textit{quantum} correction to
the classical definition of ``interval'': the complete form of the proper
distance  $s$ between two events, labeled by classical coordinates $x$ and $y$,
is given by \cite{paddy}
\begin{eqnarray}
s^2&=& \left(\,\langle\, \hat x-\hat y\,\rangle\, \right)^2 + 
       \langle\, \left(\, \hat x-\hat y\,\right)^2\,\rangle\nonumber\\
   &=& \left(\, x- y\, \right)^2 + l_0^2
\label{squantum}
\end{eqnarray}      
where $\hat x$ and $\hat y$ are quantum mechanical position operators, and
the average is defined with respect to an appropriate set of eigenstates.
Equation (\ref{squantum}) shows that the classical definition of distance
breaks down when separation becomes comparable with $l_0$ where spacetime 
itself is subjected to quantum fluctuations. The back-reaction of these
spacetime fluctuations makes the Feynman propagator ultraviolet finite.
With hindsight, it can be useful to work out the Fourier transform of
(\ref{greenl0})
\begin{eqnarray}
G\left(\, p\,\right)&=&\int \frac{\ud^4u}{(\, 2\pi\,)^2} \, e^{i\,p\cdot u}\,
G\left(\, u\,\right)\nonumber\\
&=& -\frac{1}{4\pi^2}\int_0^\infty \ud \tau\int \frac{\ud^4u}{(\, 2\pi\,)^2} 
\, e^{i\,p\cdot u}\, e^{-\tau\,(\, u^2 + l_0^2 \,)}\nonumber\\
&=& -\frac{1}{16\pi^2}\int_0^\infty \frac{\ud\tau}{\tau^2}\, e^{-\tau\, l_0^2}
e^{-p^2/4\tau\,}
\nonumber\\
&=& -\frac{l_0}{8\pi^2\sqrt{p^2}}
\int_0^\infty \frac{\ud s}{s^2}\, \exp\left[\, -\sqrt{p^2}\, \frac{l_0}{2}
\left(\, s+ \frac{1}{s}\,\right) \,\right]
\nonumber\\
&=& -\frac{l_0}{4\pi^2\sqrt{p^2}}\, K_1\left(\, l_0\sqrt{p^2}\,\right)
\label{fp}
\end{eqnarray}
where $K_1(x)$ is a Modified Bessel Function. In the UV limit one finds
\begin{equation}
G\left(\, p\,\right)\propto \frac{l_0^{1/2}}{(\, p^2\,)^{3/4}}
\exp\left(\, -l_0\sqrt{p^2}\,\right)
\label{uv}
\end{equation}

Eq. (\ref{uv}) shows as the propagator is exponentially damped at high energy
leading to finite scattering amplitudes.\\
The regular Green function $G\left(\, x-y\ ; l_0\,\right)$ can be written
as a series
\begin{equation}
G\left(\, x-y\ ; l_0\,\right)=\sum_{n=1}^\infty 
\left(\, 2\pi\, l_0\,\right)^{2(n-1)}
\left(\, G_0\left(\, x-y\,\right)\,\right)^n
\label{gl0}
\end{equation}
It is worth remarking that while each single term in the series (\ref{gl0})
is divergent in the coincidence limit, the infinite sum of all these terms
is finite \cite{del}. 
Thus, the \textit{appearance of $l_0$ is a non-perturbative effect} 
which needs at least of a partial re-summation of the standard perturbative 
expansion. Now consider
\begin{equation}
\frac{1}{n!}\left(\, G_0\left(\, x-y\,\right)\,\right)^n
=\langle\, 0\,\vert \, T\left(\, \phi^n(x)\, \phi^n(y)\,\right)\, \vert\, 0\, 
\rangle
\end{equation}
where $T$ means the time-ordering prescription, then 
guessing a suitable Lagrangian leading to (\ref{gl0}) is possible:
\begin{equation}
L= \frac{1}{2}\partial_\mu \phi\, \partial^\mu \phi -\sum_{n=1}^\infty 
\frac{1}{(n-1)\sqrt{n!}}\left(\, 2\pi\, l_0\,\right)^{n-1}\, \phi^n
\label{nopol}
\end{equation}

This simple model, even though pedagogical, is built in an ``ad hoc'' manner
to produce the effect we are looking for, that is the  replacement
(\ref{greenl0}). We can make our arguments more convincing and physically
meaningful by switching gravity on \cite{salam}.
Instead of introducing a non-polynomial interaction (\ref{nopol}) ``by hand'',
it is much more natural to couple matter field to gravity in a generally 
covariant way, e.g.
\begin{eqnarray}
L&=& \frac{1}{2}\sqrt{-g}\,
 g^{\mu\nu}\,\partial_\mu \phi\, \partial_\nu \phi +L_{EH}(\, g_{\mu\nu}\,) \\
L_{EH}&=&\frac{1}{16\pi\, G_N}\sqrt{-g}\, g^{\mu\nu}\, R_{\mu\nu}
\end{eqnarray}
where $G_N\approx L^2_{Planck}$, $g\equiv \mathrm{det}\left(\,
g_{\mu\nu}\,\right)$ and $R_{\mu\nu}$ is the Ricci tensor.
The coupling to gravity provides the needed degree of non-polynomiality
to cure the UV infinities. 
As a matter of fact, it provides more non-linearity than necessary
and one is forced to the weak-field approximation, i.e. to
expand the model around flat background spacetime according with 
$g_{\mu\nu}(x)=\eta_{\mu\nu} + \sqrt{16\pi\, G_N}\, h_{\mu\nu}(x)$, 
in order to perform calculations. Moreover, the  computations amount
to a re-summation of a particular set of Feynman graphs. Even if the result
obtained in such a way is intriguing, it cannot be considered the final
solution of the problem because it only involves a special class of 
Feynman graphs in the framework of linearized gravity.

With these results in mind, we will now switch to the 
modern formulation of quantum field theory in terms of path integral.\\
In earlier works it was shown that \cite{padma2},\cite{shanky} if the path 
integral amplitude used for the definition of
the Euclidean propagator $\exp\left[\, i\, l\left(\, x, y\,\right)\,\right]$ is 
modified so that it is invariant under the duality
transformation $l\left(\, x\ ,y\,\right)\to l_0^2/l\left(\, x\ ,y\,\right)$,
then:\\
i) the propagator becomes UV finite, and\\ 
ii)  $l_0$ represents a residual, or zero-point length, somehow related to 
     spacetime fluctuations at short distance.\\
Thus, the basic information is that quantum
gravity is an \textit{effective cure} for UV infinities. But, what is
quantum gravity beyond the weak-field approximation?\\
In the eighties the description of short distance spacetime structure
took a completely new direction 
with the \textit{revolutionary} re-appearance of string theory. Born in the
sixties as a possible candidate to describe strong interactions among
extended hadronic objects \cite{rebbimand}, string theory was dismissed after the discovery
that a relativistic invariant quantum formulation was self-consistent only
in higher dimensional spacetime.
Early observation that the first excited level of a closed string contains
a massless, spin-two state, remarkably similar to the expected quantum
of the gravitational field was not taken seriously for almost twenty years.
The mind barrier to accept the concept of higher dimensional spacetime
was slowly demolished by the pointless attempts to build Grand Unified,
eventually Supersymmetric, gauge models in four spacetime dimensions.
Finally, in a couple of seminal papers, Green and Schwartz showed that
a string theory with internal symmetry group $SO(32)$ or $E_8\otimes E_8$ is
gauge and gravitational anomaly-free \cite{greschwa}. Therefore, it provides a 
self-consistent
framework where all fundamental interactions, including gravity, can be
quantized in a physical meaningful way.
String theory has both these features (~non-polynomiality as well as finite length 
scale~) built into it in a natural fashion. This suggests that one should be 
able to obtain similar results in standard string theory.
In this paper we  show that this is indeed the case: the 
zero-point length can be reproduced using $T$-duality of the string theory, and
the leading order corrections to 
the propagator, defined in a specific but consistent way, is the same as that 
obtained from the principle of path integral duality used in  \cite{padma2}.

In this paper we are going to put in a single consistent framework apparently
unrelated pieces of information, i.e. zero-point length, extra-dimensions,  
string $T$-duality.
 More in details, we are going to introduce a $4D$ non-trivial vacuum where
 topologically non-trivial fluctuations take place. These kind of virtual
 processes are sensitive both to the presence of extra-dimensions and to
 the string excitation spectrum. It follows that
 zero-point length can be seen as a four dimensional 
``virtual memory'' of compact extra-dimensions length scale. Furthermore, such 
a scale turns out to be determined by $T$-duality inherited from the underlying 
fundamental string theory. 
From a low energy perspective short distance infinities are cut off by
a minimal length, which is proportional to the square root of the string
slope defined as $\sqrt{\alpha^\prime}$. Thus, we bridge the gap between the
ultra-relativistic, ten-dimensional, string domain and the four-dimensional
low energy arena of point-particle quantum field theory.


\section{Effective $4D$ propagator in the Kaluza-Klein vacuum}

Higher dimensional spacetime and compact extra-dimensions were introduced
in theoretical physics much before the advent of string theory. 
In 1921 the Polish mathematician Theodor Kaluza published one of the first
attempt to unify gravity and electromagnetism in a single geometrical 
framework \cite{kaluza}.
The crux of Kaluza model was the shocking suggestion that spacetime is $4+1$
dimensional, rather than $3+1$ dimensional. In order to justify that
physical quantities appear to vary only with respect to the usual $3+1$ 
coordinates, Kaluza introduced as an ansatz that the five dimensional metric
was independent from the fifth coordinate, $x^5$.\\
Kaluza's original proposal was refined few years later by Oscar Klein
\cite{klein}:
rather than assuming higher dimensional metric to be independent from $x^5$,
Klein suggested the five dimensional spacetime to be periodic in the new extra
coordinate. Thus, $5D$ metric tensor can be Fourier expanded
\begin{equation}
g_{MN}\left(\, x^\mu\ , x^5\,\right)=\sum_{n=-\infty}^{\infty} 
g_{MN}^{(n)}\left(\, x^\mu\,\right)\exp\left(\, i \, n \frac{x^5}{R}\,\right) 
\end{equation}
(~Greek indices run from $0$ to $3$ while Latin ones count the fifth component 
too~) 
where the fifth dimension is a circle of radius $R$, and $0\le x^5 \le 2\pi R$.
By assuming $R$ to be very small, may be few orders of magnitude larger
than the Planck length, then we cannot observe any physical effect of the
extra-dimension in everyday life. Only probes with energy comparable to
the Planck energy can ``feel'' the presence of the extra-dimension. 
Let us note that the ``zero-mode'' in the Fourier series is $x^5$ independent
and can be identified with the $4D$ metric tensor.  

The Dual Models of hadrons, introduced in the sixties to account for the
high energy behavior of strongly interacting particles, required the presence
of extra-dimensions in a completely different, and apparently unrelated,
framework. The linearly fitting hadronic resonances in the plane mass$^2$-spin
pointed out the existence of a universal  one-dimensional structure underlying
their phenomenological Regge behavior \cite{collins}. Rather than dimensionless 
massive
points, strongly interacting particles behaved as one-dimensional, rotating,
relativistic ``strings'', i.e. one-dimensional vibrating filaments of pure
energy. But, this groundbreaking idea crashed against an apparently insolvable
contradiction: Lorentz invariance was preserved at the quantum level only if
spacetime was either $26$ or $10$ dimensional! \\
In the ordinary four dimensional
spacetime a quantum theory which describes hadrons as one-dimensional objects
cannot be quantized in any way compatible with Special Relativity.
This objection was by-passed when in the eighties string theory was back
on the spot again, no more as a theory of strongly interacting particles
but as the best candidate to the role of unified theory including gravity.\\
It worth to notice that Dual Models provide scattering amplitude with
a much better ultraviolet behavior than any renormalizable, or even
super-renormalizable\-ble, quantum field theory \cite{collins}. Part of the price 
to pay to cure high energy infinities is that spacetime must be higher 
dimensional.\\
Before going to consider full string theory, let us start from a simpler, 
probably more familiar, setting where the interplay between extra-dimensions
and zero-point length shows up. Let us consider a \textit{free} quantum
mechanical, point-particle  in a Kaluza-Klein background geometry. For the sake
of simplicity, we consider a $4+1$ dimensional, flat spacetime where the
fifth space-like dimension is a \textit{circle} of radius $R$, i.e. 
$0\le x^5\le 2\pi\, R$, the generalization to an higher dimensional torus
is straightforward.\\
To capture the relation between  zero-point length
and the radius of extra-dimension, we need to determine the particle 
propagator.
One of the most effective way to compute the propagation amplitude is through
a ``sum over paths''. Going more in details, the quantum mechanical amplitude for a
particle of mass $m_0$ to ``jump'' from an initial position $x^M\equiv
\left(\,x^\mu\ , x^5 \,\right)$ to a final position $y^M\equiv
\left(\,y^\mu\ , y^5 \,\right)$ can be formally written as a sum over all
possible paths starting from $x^M$ and ending $y^M$ in a time lapse $T$.
Each path is weighted by a factor $\exp\left(\, i \, S/\hbar\,\right)$, where
$S$ is the classical action associated to the path and $\hbar$ is the Planck
constant \cite{feyn}. To keep the notation as light as possible we set $\hbar=1$. As far as
the action of the test particle is concerned, we have a certain freedom in the
choice of its explicit form. Again, in order to keep calculations as simple as 
possible, we take
\begin{equation}
S\equiv \int_0^T \ud \tau\,\left[\, p_M\, \dot z^M - \frac{1}{2\mu_0^2}\,\left(\,
 p_M\, p^M\, + m_0^2\,\right)\,\right]\label{s0}
\end{equation}
where, $z^M=z^M(\,\tau\,)$, $z^M(\,0\,)=x^M$ and $z^M(\,T\,)=y^M$. The ``dot''
represents the derivative with respect the affine parameter $\tau$
(chosen dimensionless) and
$p_M$ is the particle five-momentum. The action (\ref{s0}) is one among several
equivalent functionals encoding classical dynamics of a relativistic
point-particle of mass $m_0$. The chosen form  corresponds
to a Lagrangian which is quadratic in the five-velocity $\dot z^M$. It is
important to remark that $\mu_0$ is a unit mass scale keeping correct physical
dimensions while $m_0$ is the physical mass of the particle.
In the point-particle case $\mu_0$ is an arbitrary mass scale, on the
other hand we shall see that string theory has its own mass
scale determined by the string tension.
Even if we do not start from the (~non-linear~) reparametrization
invariant form of the action written in terms of the trajectory proper-length, 
we shall obtain the relativistic Feynman propagator by averaging over the
unobservable lapse of time $T$.\\
Our main goal is to link the presence of $l_0$ in the four
dimensional propagator to some effect induced by the
existence of a non directly observable fifth dimension. Since we are
looking for a four dimensional propagator, we will project the five
dimensional results into the usual $3+1$ dimensions by selecting
all possible paths connecting two distinct points
(~$x^\mu$ and $y^\mu$~) in four dimensions  while describing 
closed loops along $x^5$. At this point, it is worth to remind that in the
path integral language virtual fluctuations are described in terms of
closed histories starting and ending at the same point \cite{kleinert}. This allows to
interpret the five dimensional propagator projected into the
four dimensional spacetime  as a propagator  of a particle in a
non trivial vacuum, i.e. in a vacuum populated by the virtual
fluctuations along $x^5$.
Moreover, as the fifth dimension is a circle, closed paths can be
classified in terms of the number $n$ of times they wind around
$x^5$, this will make $l_0$ appear in a natural way. Thus, we can  write
\begin{eqnarray}
K\left[ x^\mu- y^\mu , 0 \,  ; T \,  \right]=N \!\!\!\!
\sum_{n=-\infty}^\infty
\int_{z(0)=x}^{z(T)=y}
\oint_{z^5(0)=x^5}^{z^5(T)= x^5+ n\,l_0} \!\!\!\!\!\!\!\!\!\! 
\left[\,Dz^\mu\,\right] \left[\,Dp_\mu\,\right]
\left[\, Dz^5\,\right]
 \left[\,Dp_5\,\right]\times &&\nonumber\\
\exp\left(\, i\int_0^T \ud \tau\left[\,p_\mu\dot z^\mu +p_5\dot z^5
-\frac{1}{2\mu_0^2}\left(\, p_\mu p^\mu +  p_5 p^5 + m_0^2\,\right)\,\right]
\,\right)&&
 \label{kernel}
\end{eqnarray}   
where $l_0\equiv 2\pi\, R$ and
$N$ is a normalization factor to be determined later on. 
Inspection of (\ref{kernel}) shows that the  kernel can be written as
\begin{eqnarray}
&& K\left[\, x^\mu-y^\mu\ , 0\ ; T\,\right]=N\, K\left[\, x^\mu-y^\mu\ ;
T\,\right] \times \nonumber\\
&& \sum_{n=-\infty}^\infty
\oint_{z^5(0)=x^5}^{z^5(T)= x^5+ n\,l_0}\left[\, Dz^5\,\right]
 \left[\,Dp_5\,\right]\,
 \exp\left(\, i\int_0^T \ud\tau\left[\, p_5\,\dot z^5
-\frac{1}{2\mu_0^2}\, p_5\, p^5\,\right]\,\right)\nonumber\\
&&
 \end{eqnarray}   
where 
\begin{equation}
K\left[\, x^\mu-y^\mu\ ;  T\,\right]=\frac{1}{(2\pi)^4} \left(\, 
\frac{\pi\,\mu_0^2}{i \, T}\,\right)^2
\, \exp \left[\, i\frac{\mu_0^2}{2T}
\, \left(\, x-y\,\right)^2-i \frac{m_0^2}{2\mu_0^2}\, T\, \right]
\end{equation}
is the standard  four dimensional amplitude.\\
The contribution from the fifth-component of the paths  can be obtained as
follows:
\begin{equation} 
\int_0^T p_5 \,\dot z^5 \, \ud\tau 
= \left[\,p_5\, z^5\,\right]_0^T
-\int_0^T \ud \tau \, z^5\, \dot{p}_5\label{p5}
  \end{equation} 
By taking into account the (~functional~) representation of the Dirac
delta-function
\begin{equation}  
\oint\left[\,Dz^5\,\right]\exp\left(-i\int _0^T \ud\tau\,
z^5\,\dot p_5 \,\right)=\delta\left[\,\dot p_5\,\right] 
\label{x5}
\end{equation}
we get:
\begin{eqnarray} 
\oint_{z^5(0)=x^5}^{z^5(T)= x^5+ n\,l_0} \!\!\!\!\!\!\!\!\!\!\!\! 
\left[ Dz^5 \right]
 \left[ Dp_5 \right]\exp\left[\, i \!\! \int_{x^5}^{x^5 +n\,l_0 }
  \!\!\!\!\!\!\!\!\!\! \ud\left( p_5\, z^5 \right)- i \!\! \int_0^T \ud\tau \,
  \left( z^5 \, \dot p_5 
+\frac{1}{2\mu_0^2} p_5\, p^5 \right) \right] && \nonumber\\
 =\int \left[\,Dp_5\,\right]\delta\left[\,\dot p_5\,\right]
 \, \exp{i\,\left[\, p_5\, z_5\,\right]^{x^5+ nl_0}_{x^5}}\,\exp\left(\,
 -\frac{i}{2\mu_0^2} \int_0^T \ud\tau \,p_5 \, p^5\,\right) &&
\label{xp5}
\end{eqnarray} 

The Dirac delta shows that $p_5$ is \textit{constant} (~$\tau$-independent~)
over all possible trajectories, and makes 
the  functional integral over $p_5$ trajectories
to collapse into an ordinary integral over the constant values of $p_5$ \cite{ej}.
Furthermore, boundary condition at the trajectory end-point determines
the \textit{discrete spectrum} of $p_5$ as follows
\begin{eqnarray}
z^5(\, \tau\, )&=&\frac{p_5}{\mu_0^2}\tau + x^5\\
z^5(\, T\, )&=&x^5 + n\, l_0 \longrightarrow p_5 = \frac{\mu_0^2}{T}\, n\, l_0
\end{eqnarray} 
giving for equation. (\ref{xp5})
\begin{eqnarray} 
 &\!\!\!\!\!\!&\!\!\!\!\!\! \int\left[\,Dp_5\,\right]
 \delta\left[\,\dot p_5\,\right]
  \exp{i\left[\, p_5\, z^5\,\right]_{x^5}^{x^5 +n\, l_0}}\exp\left(\,
 -\frac{i}{2\mu_0^2} \int_0^T \ud\tau \,p_5 \, p^5\,\right)\nonumber\\
 && = \int \frac{\ud p_5}{2\pi}\delta\left(\,p_5 - \frac{\mu_0^2}{T}\,n\, 
l_0\,\right)\,
  \exp \left(i p_5\, n\, l_0\, -\frac{i T}{2\mu_0^2}\,
 p_5 \, p^5\,\right) \nonumber\\
 &&=  \exp\left[\,\frac{i\, \mu_0^2 }{2T}\, n^2\,l_0^2\,\right]\label{k5}
 \end{eqnarray} 
Thus, we get 
\begin{eqnarray}
 K\left[\, x-y\ ; T\,\right]&=&
\frac{N}{(2\pi)^4}\, 
\, \left(\, \frac{\pi\, \mu_0^2}{i T} \,\right)^2\times\nonumber\\
&&\sum_{n=-\infty}^\infty\, \exp \left\{\, i\frac{\mu_0^2}{2T}
\left[\, \left(\, x-y\,\right)^2  + 
n^2\,l_0^2\,\right] -i\frac{m_0^2}{2\mu_0^2}\, T\,\right\}
\label{kred}
\end{eqnarray}

The amplitude (\ref{kred}) describes the propagation of a particle of mass
$m_0$ through a non-trivial vacuum: the ``memory'' of quantum fluctuations 
of the paths along $x^5$ shows up as a residual length, $n\, l_0$, of the
paths connecting $x$ to $y$.\\
The modified amplitude (\ref{kred})
allows to discuss the problem of short distance divergences. In fact,
\begin{equation}
K\left[\, x^\mu-y^\nu\ ; T\to 0\,\right]\approx \sum_{n=-\infty}^\infty\,
\delta\left[\, \left(\, x-y\,\right)^2 + n^2\,l_0^2\, \right]
\label{kshort}
\end{equation}
While the Dirac delta is always zero for any $n\ne 0$, the zero-mode
is singular in the coincidence limit $x\to y$. Indeed the $n=0$ mode 
corresponds
to the standard propagator in four dimensional spacetime with all its
ultraviolet divergences. We recall that $n=0$ term corresponds to virtual
fluctuations described in terms of closed path along $x^5$ which are
contractible to a point. As such, they can be shrunk to arbitrary short
length and cannot really ``feel'' the presence of 
the compact extra-dimension. The current wisdom is to keep anyway this family 
of paths and introduce an appropriate
regularization method to deal with short distance infinities.
On the other hand, if we believe that  a minimal length is embedded into the
spacetime fabric the short time asymptotic form of $K\left(\, x^\mu-y^\nu\ ;
T\,\right)$ cannot contain any divergent term as $x\to y$. Self-consistency
requires we drop the zero-mode term in (\ref{kred}):
\begin{equation}
K\left[ x^\mu-y^\nu ; T \right]=
\frac{2\,N}{(2\pi)^4} 
\left( \frac{\pi\, \mu_0^2}{i T} \right)^2
\sum_{n=1}^\infty\, \exp \left\{ i\frac{\mu_0^2}{2T}
\left[ \left( x-y \right)^2  + n^2\,l_0^2 \right] -i\frac{m_0^2}{2\mu_0^2}
\, T \right\}
\label{kreg}
\end{equation}

On the other hand, we cannot give up the standard four-dimensional kernel:
once the zero-mode is dropped out we can recover it from the limit $l_0\to 0$
\begin{equation}
\lim _{l_0\to 0 }\, K\left[\, x^\mu-y^\nu\ ; T\,\right]=
\frac{2\,N}{(2\pi)^4} 
 \left( \frac{\pi\,\mu_0^2}{i T} \right)^2
 \zeta(0)\, \exp \left\{\, i\frac{\mu_0^2}{2T}\,
 \left(\, x-y\,\right)^2 -i\frac{m_0^2}{2\mu_0^2}\, T\,\right\}
\label{drop}
\end{equation}
where $\zeta(0)$ is the Riemann zeta-function in zero:
\begin{equation}
\zeta(0)\equiv \sum_{n=1}^\infty\, \frac{1}{n^0} = -\frac{1}{2}
\end{equation}
and the canonical form of $K\left[\, x^\mu-y^\nu\ ; T\,\right]$ is recovered
by choosing in appropriate way the normalization constant.\\
Therefore, the regularized form of the
kernel is obtained by dropping the $n=0$ mode and re-writing (\ref{kred})
as 
\begin{equation}
K_{reg}\left[\, x-y\ ; T\,\right]=
\frac{\mu_0^4}{2\pi^2T^2}
\, \sum_{n=1}^\infty\exp \left\{\, i\frac{\mu_0^2}{2T}
\left[\, \left(\,x-y\,\right)^2 +n^2l_0^2\,\right] -i\frac{m_0^2}{2\mu_0^2}\, T 
\,\right\}
\label{Kreg}
\end{equation}

It is important to remark that 
by dropping out the zero-mode we do not loose any information since
the standard $4D$ particle propagator is recovered anyway in the limit 
$l_0 \to 0$,  i.e. the $n=0$ mode is traded for the limit $l_0\to 0$. 
It is worth stressing that, so far, the prescription of dropping $n=0$
mode corresponds to a particular choice of \textit{regularization} of the
kernel. At this level the kernel can be considered as deriving from
a Kaluza-Klein Quantum Field Theory with the new input that the length
$l_0$ of the compact dimension acts as a universal regulator of the 
short distance behavior in  $4D$ spacetime.
Short distance finiteness of measurable quantities is assured by the presence 
of $l_0\ne 0$ which, however, remains an arbitrary parameter. The improvement
we get up to this stage is that
we do not need to introduce ``by hand''  an ultraviolet cut-off, or
a  subtraction point,  because the regulator parameter is present from the 
very beginning in the spacetime geometry the test particle is propagating 
through. 

We can show explicitly how $l_0$ regularize UV divergences
by working out the  Green function from (\ref{Kreg}) 
\begin{eqnarray}
G\left(\, x-y\,\right)&\!\!\!\!\!\! \equiv&\!\!\!\!\!\! 
-\frac{\pi^2}{\mu_0^2}\int_0^\infty \ud T\, K_{reg}\, \left[\, x-y\ ; T\,\right]
\nonumber\\
&\!\!\!\!\!\!=&\!\!\!\!\!\! -\frac{\mu_0^2}{2\pi}
\sum_{ n=1}^\infty
\int_0^\infty  \frac{ \ud T}{T^2}\exp\left( -\frac{i\,T}{2\mu_0^2} 
m_0^2 \right)\,
\exp \left\{\frac{i \, \mu_0^2  }{2T}
\left[\, \left(\,x-y\,\right)^2 +n^2l_0^2\,\right] \right\}\nonumber\\
\label{greg}
\end{eqnarray}
In order to study the short-distance behavior of (\ref{greg}) it is enough to
consider the leading $n=1$ term
\begin{equation}
G\left(\, x-y\,\right)\approx -\frac{\mu_0^2}{2}
\int_0^\infty  \frac{ \ud T}{T^2}\exp\left(\, -\frac{i\,T}{2\mu_0^2} 
m_0^2\,\right)\,
\exp \left\{\,\frac{i\, \mu_0^2 }{2T}
\left[\, \left(\,x-y\,\right)^2 +l_0^2\,\right] \,\right\}
\label{gl}
\end{equation}
The Fourier transform of $G\left(\, x-y\,\right)$ reads
\begin{eqnarray}
G\left(\, p\,\right)&\equiv& \int \frac{\ud^4u}{(2\pi)^2} \, 
e^{i\, p\, u} 
\,G\left(\, u\,\right) \nonumber\\
&=&\int_0^\infty \ud t \exp\left[\,- it \,\left(\, p^2 + m^2_0 \,\right)
+\frac{i l_0^2}{4 t}\,\right] , \quad t\equiv\frac{T}{2\mu_0^2}
\end{eqnarray}
By re-absorbing $\mu_0^2$ in the definition of the Schwinger time parameter
$t$ one eliminates the unit mass scale leaving to the physical mass
$m_0$ the role to determine the position of the pole in the Feynman propagator.
Indeed, it is straightforward to recover the standard Feynman propagator in the
limit $l_0\to 0$:
\begin{equation}
\lim_{l_0\to 0}\, G\left(\, p\,\right)=\int_0^\infty \ud t \exp\left[\,- it
\,\left(\, p^2 + m^2_0 \,\right)\,\right]=\frac{-i}{p^2 + m^2_0}
\end{equation}
while, by keeping $l_0\ne 0$ we get
\begin{equation}
G\left(\, p\,\right)= \frac{l_0}{\sqrt{ p^2 + m^2_0}}\, K_1\left(\, 
l_0\, \sqrt{ p^2 + m^2_0} \right)
\label{gpreg}
\end{equation}
where $K_1(x)$ is a modified Bessel function
and $p^2$ is now the squared momentum in Euclidean space. 
The high energy behavior
of (\ref{gpreg}) is recovered by taking asymptotic form of $K_1(x)$ for
large Euclidean momentum. In this regime we find
\begin{equation}
G\left(\, p\,\right)\approx \frac{\sqrt{l_0}}{\left(\, p^2\,\right)^{3/4}}
\exp\left( -l_0\, \sqrt{p^2} \, \right)
\label{urel}
\end{equation}
Equation.(\ref{urel}) shows that the Feynman propagator is exponentially damped
at ultra-relativistic energy (~$ p^2 >> m_0^2 $, $ p^2 >> l_0^{-2} $~)
thanks to the presence of $l_0$. This natural
exponential cut-off makes UV finite any local field theory scattering 
amplitude.

In higher dimensional Quantum Field Theory one expects
$l_0 \! = \! O\left( 10 L_{Planck}\right)$, but the actual value of the
extra-dimension(s) length scale remains beyond the predictive capability
of the models. Despite many remarkable attempts, as to use Casimir effect
to determine $l_0$ \cite{candelas}, the intrinsic
non-perturbative nature of the problem crashes against the perturbative
computational techniques available up to now, e.g. loop-expansion of the
vacuum energy density. Determining $l_0$ in a
non-perturbative way needs string theory as we shall see in the next 
sections.

\section{A ``preview'' of string dynamics and $T$-duality}
 
In this section we would like to give the reader the main ideas
and results of our approach without going into the technical
details which are discussed in the next section.

A fundamental string is a physical object characterized by three basic
features:
\begin{enumerate}
\item[1)] it is spatially extended in one dimension;
\item[2)] it is a relativistic object;
\item[3)] it can accommodate supersymmetry.
\end{enumerate}
Item $3)$ is not relevant for our discussion, thus to keep the discussion
as simple as possible we shall consider a closed bosonic string. The extension
to a full supersymmetric string requires only changing the critical
dimension in the final results.\\
Items $1)$ and $2)$ imply that the string is dynamically equivalent
to an infinite set of relativistic harmonic oscillators \cite{book}. Lorentz invariance
implies that not all the components of every harmonic oscillator amplitude
correspond to physical oscillations. Accordingly, a proper gauge fixing
procedure will be required to select and quantize physical degrees of
freedom alone. All that will be described in details in the next section.
Now, let us simply ``guess'' the string spectrum on the basis of the following
arguments.
One can always look at the string as a ``mechanical'' extended
object. From this point of view, the motion of the string can always be
decomposed in terms of the motion of its center of mass and the relative
motion of its constituents (~with respect the center of mass~):
\begin{center}
string coordinates $=$ center of mass coordinates $+$ relative  coordinates
\end{center}
The motion of the string as a whole is described by the motion of a
fictitious point-like object with mass equal to the total mass of
the string itself. Item 1) requires we add the contribution from the
``internal energy'', i.e. both excitation and zero-point energy of
any string point. By putting all together we find
\begin{eqnarray}
\hbox{string energy}&=&\hbox{center of mass energy}+\nonumber\\
&&\hbox{oscillator spectrum} +\hbox{total zero-point energy}
\label{spectr}
\end{eqnarray}
Item 2), i.e. Lorentz invariance,  has the drawback to over-count
degrees of freedom by including timelike and longitudinal oscillators.
In a spacetime with $d$ spacelike dimensions only $d-1$  transverse 
oscillation modes  correspond to physical motions and contribute
to the physical  spectrum of the system. With this caveat we can write down
the expected form of the energy spectrum
\begin{equation}
E^2 = {\vec P}_0^{\, 2} + 2\pi\rho\left[\, 
N +\tilde{N}\, -\frac{d-1}{12}\,\right]
\label{mass}
\end{equation}
where $\vec{P}_0$ is the transverse part of the
 string center of mass momentum, $\rho$ is the string tension,
$N$ and $\tilde{N}$ are the occupation numbers counting the number
of harmonic excitations on each level and $2\pi\rho(d-1)/12$ is the
regularized contribution from the  zero-point energy. The tension is
often written as
\begin{equation}
\rho\equiv \frac{1}{2\pi\alpha^\prime}
\label{regge}
\end{equation}

For historical reasons
the constant $\alpha^\prime$ is known as the \textit{Regge slope} \cite{collins}.
\\
It is immediate to recover from (\ref{mass}) the corresponding
mass spectrum, that is
\begin{equation}
M^2= 2\pi\rho\left[\, 
N +\tilde{N}\, -\frac{d-1}{12}\,\right]
\label{0mass}
\end{equation}

Now, along each compact dimension the momentum is ``quantized'' according to 
the
following semi-classical argument: to a classical momentum $P^i_0$ one 
associate a De
Broglie wavelength $\lambda_i=1/ (2\pi P^i_0)$; only an integer number of 
wavelengths can fit into the compact dimension. For the sake of simplicity
we take all the compact extra-dimensions to be circles of radius $R_i$,
 hence we have $2 \pi R_i=n_i \lambda _i$ and
\begin{equation}
P^i_0=\frac{n^i}{R_i}\ ,\qquad n_i=0\ ,\pm 1\ , \pm 2\ ,\dots
\label{db}
\end{equation}
where the index $i$ labels transverse, compact, space-like dimensions.
Let us remark that this is not the argument we advocated in the path-integral
formulation of the KK-particle propagator. Indeed, in Feynman integral we sum
over \textit{trajectories} while Equation. (\ref{db}) assumes the wave-particle
duality relation. The final result for the propagator is the same in both
cases, but we considered more appropriate to avoid wave-like assumptions inside
the functional integral over trajectories. However,  advocating particle-wave
duality is a short-cut to the wanted result. So far, we have only recovered
the same kind of condition we found in Kaluza-Klein models.  The novelty
is that closed strings have the unique property to wrap around compact 
spatial dimensions. 
In the case of a single compact dimension, the string 
mass spectrum is labeled, in addition to $N$, $\tilde{N}$ and $n$,
by a new quantum  number $w$ which counts 
the string winding around compact dimension. It is easy to understand
how these topological configurations contribute to the mass. The string
\textit{tension} represents the energy per unit
length of any static classical configuration. If the string wraps $w^i$
times around $x^i$ its energy will increase according with:
\begin{eqnarray}
 M_w &\equiv& \left(\, \hbox{tension}\,\right)\times \left(\,
 \hbox{total length}\,\right)\nonumber\\
& =&\rho \, w_i\, 2\pi R_i
\end{eqnarray}
Thus, equations (\ref{mass}),  (\ref{0mass}) read:
\begin{eqnarray} 
E^2 &=&\sum_i\frac{n^2_i}{R^2_i} + 4\pi^2 \rho^2 \,\sum_i w^2_i\, R^2_i +
    \hbox{quantum oscillators}\, \label{M}\\
M^2 &=&4\pi^2 \rho^2 \,\sum_i w^2_i\, R^2_i +
    \hbox{quantum oscillators}\, \label{N}
\end{eqnarray}
The most important property of the spectrum (\ref{M}) is its invariance
under, so-called, $T$-duality which is defined as a transformation  relating 
the compactification radius and its inverse while, at the same time,
exchanging quantum numbers $n$ and $w$ as:
\begin{eqnarray}
&& n_i\longleftrightarrow w_i \label{T1}\\
&& R_i\longleftrightarrow \frac{1}{2\pi \rho\,  R_i}\label{T2}
\end{eqnarray}
The meaning of such a symmetry is that \textit{ a closed string cannot
distinguish KK modes on a circle of radius $R_i$ from winding modes around a
circle of radius $\tilde R_i = \alpha^\prime/R_i$.} The distinction is a pure
conventional matter. Physical quantities as the energy spectrum are
invariant under (\ref{T1}), (\ref{T2}). \\
This symmetry has a very important consequence:
there exists a  \textit{Minimum} radius
of compactification which is invariant under (\ref{T1}), (\ref{T2}).
 This is the minimal radius a closed string can wrap around.
It is called the \textit{Self-Dual radius}:
\begin{equation}
R^\ast_i =\frac{1}{\sqrt{2\pi \rho}}= \sqrt{\alpha^\prime}
\label{rmin}
\end{equation}

Thus, the quantum mechanical concept that probing
shorter distances requires more and more energetic test objects breaks down
when applied to a string whose minimum size is given by (\ref{rmin}).\\
To establish a connection between self-dual radius and zero-point length
we have to compute an effective Feynman propagator following the same strategy 
we discussed in the previous section. That is, we are going to investigate
how the center of mass of the string propagates in a non-trivial
four dimensional vacuum which is populated by all kind of string
virtual fluctuations. For the sake of simplicity let us stick
to the $4+1$ dimensional case and  momentarily drop
the oscillator modes, their presence and the critical spacetime dimension  
are not relevant for the present discussion. The complete form of the propagator 
will be given in the next section.

In the  Schwinger proper time representation of the Feynman propagator
the mass $m_0$ in the weight factor   $\exp \left( -im^2_0\,\tau\right)$ 
is the physical mass of the particle, or
the pole in the momentum space propagator \cite{schwi}. In order to relate our 
previous calculations
to string theory we shall choose the weight factor as the string mass 
contribution coming from Kaluza-Klein and winding modes. The latter modes are
intrinsically extended object configurations which are absent in the spectrum 
of any
point-like object. Their inclusion is the crucial link to string theory.
In order to avoid any confusion, we remark that $n$ counts the encircling of 
the path integral string center of mass histories around compact dimension, 
while $w$ refers to
the  wrappings of the string itself. To account for possible low-energy
non-perturbative effects, e.g. spontaneous symmetry breaking,
vacuum condensates, etc., we add to (\ref{M}) an, eventual, 
non-stringy contribution $m_0$. 
Thus, we expect the string induced effective Green function  of the form
\begin{eqnarray}
G_{reg}\left(\,x-y\,\right)&\approx& 
\sum_{w,n=1}^\infty
\int_0^\infty \frac{\ud t}{t^2} \, \exp \left[-i\,t \left( m_0^2 + \frac{w^2 \,
l_0^2}{\left(2 \pi \alpha^\prime \right)^2} \right) \right] \times \nonumber\\
&& \exp \left\{ \frac{i}{4 t} \left[ \left(\,x-y\,\right)^2 +n^2l_0^2\,\right] \,\right\}
\label{greg2}
\end{eqnarray}
The various summations take care of both Kaluza-Klein and winding quantum
fluctuations. We use the symbol $\approx$ as a reminder for dropping
the string oscillator modes and the overall normalization constant. Furthermore, 
we have eliminated both zero modes $n=0$ and $w=0$ according with
our previous discussion. Winding number $w=0$ describes an
un-wrapped string. As far as no oscillator modes are excited it cannot be
distinguished from its own center of mass which has been already taken in account.
\\
Integration over $t$ variable leads to a modified Bessel function $K_{-1}(z)$
(which is equal to $K_1(z)$).
Thus, the Green function is given by:
\begin{equation}
G_{reg}\left(x-y\right)\approx 
\sum_{ w , n=1}^\infty
\frac{M_w}{\sqrt{\left( x-y \right)^2 +n^2\,l_0^2}}
 \, K_{1}\left(M_w\sqrt{\left(x-y \right)^2+n^2\,l_0^2 } \right)
\label{final}
\end{equation}
with
\begin{equation}
M^2_w\equiv m_0^2+ \frac{w^2 \,l_0^2}
{\left(2 \pi\alpha^\prime \right)^2} 
\label{mspectrum}
\end{equation}

Finally,
$T$-duality naturally selects an invariant value of the zero-point length
$l_0=2 \pi \sqrt{\alpha^\prime}$ as the unique compactification radius
which is self-dual. Thus, $T$-duality is
instrumental to connect an ``a priori'' arbitrary parameter, $l_0$, or $R$, to
the string length scale $\sqrt{\alpha^\prime}$. In this way string theory
provides a fundamental short distance cut-off for low energy quantum field 
theory \cite{padma1} in the form of a residual zero-point length 
\cite{padma2}. The leading term in (\ref{final}) is the one corresponding to 
$n=w=1$:
\begin{equation}
 G_{reg}\left(\,x-y\,\right)\approx
\frac{m}{\sqrt{\left(x-y\,\right)^2 +l_0^2}}
 \, K_1\left(m \sqrt{\left(x-y\,\right)^2 +\,l_0^2}\,\right)
\end{equation}
where
\begin{equation}
m^2 \equiv m_0^2 + \frac{l_0^2}{\left(2\pi \alpha^\prime \right)^2}
\end{equation}
The Fourier transform of this term gives the (euclidean) propagator in 
 momentum space, which is
\begin{equation}
 G_{reg}\left(\,p\,\right)\approx
\frac{l_0 }{\sqrt{p^2 + m^2}} K_1\left(\, l_0\sqrt{\,p^2+m^2}\,\right)  
\label{tram}
\end{equation}
Equation (\ref{tram}) is the result found earlier in  \cite{padma2}.


\subsection{Strings as infinite sets of harmonic oscillators}

The evolution of a one-dimensional object sweeps a two dimensional
\textit{world-surface} embedded into \textit{target} spacetime. 
The world-surface of a closed relativistic string
can be described by introducing an evolution parameter
$\tau$ taking values in the real interval ${\mathcal{I}}=
\left[\, 0\ , T\,\right]$
and letting $X$ be a map from ${\mathcal{I}}\times S^1$ into target 
spacetime.
The image of ${\mathcal{I}}\times S^1$ under the embedding $X$ describes in 
a Lorentz covariant way string evolution in spacetime. Each point on the 
world
surface is identified by set of coordinates $x^\mu(\, \tau\ ,\sigma\,)$ 
where $\mu=
0\ , 1\ ,\dots d$ is a target spacetime index. $d$ is the number of spatial
dimensions. $\sigma$ is a spacelike coordinate 
in the range $ 0\le \sigma \le \sigma_0$.\\
The action for the string coordinates  $x^\mu$ is taken in analogy
with the action of a relativistic point-particle by replacing the length of the
particle world-line with the area of the string world-surface:
\begin{eqnarray}
S[\, x\,] &=& -\hbox{tension}\times \hbox{world-sheet area}\nonumber\\
&=&-\frac{1}{2\pi\alpha^\prime}\int_0^T \ud\tau\int_0^{\sigma_0} \ud\sigma
\sqrt{-det(\,\gamma \,) }
\label{action}
\end{eqnarray}
where, 
$\gamma_{mn}= \eta_{\mu\nu}\,\partial_m\, x^\mu\, \partial_n\, x^\nu $
is the \textit{induced metric} on the string world-surface. The proportionality
constant is the string tension $\rho$ as defined in (\ref{regge}).
For convenience one can rescale $\sigma\rightarrow \pi\sigma/\sigma_0$.\\
The Nambu-Goto action (\ref{action}) is non-polynomial and contains
the single dimensional constant $\alpha^\prime$ with length squared dimension
in natural units \cite{nambu}. Thus, it displays all the needed features we 
discussed in the introduction to obtain a regular short distance behavior.\\
Beside its clear geometric meaning, the action (\ref{action}) is 
manifestly invariant under
local coordinate transformations $\left(\, \tau\ ,\sigma\,\right)\rightarrow
\left(\, \tau^\prime\left(\, \tau\ ,\sigma\,\right) \ ,
\sigma^\prime \left(\, \tau\ ,\sigma\,\right) \,\right) $. \\
A form of $S$ more appropriate to perform path integral calculations
can be obtained by introducing an auxiliary world-sheet metric $g_{mn}$,
not to be confused with the induced metric $\gamma_{mn}$, or the
target spacetime metric $\eta_{\mu\nu}$. Then, we can write Nambu-Goto action 
as \cite{polya}
\begin{equation}
S=-\frac{1}{2\pi\alpha^\prime}\int_0^T \ud\tau\int_0^\pi \ud\sigma
\sqrt{-det(\,g\,) }\, g^{mn}\, \eta_{\mu\nu}\,
\partial_m\, x^\mu\,\partial_n\, x^\nu \,  \label{spol} 
\end{equation}

The Polyakov action (\ref{spol}) provides an alternative description of the
string in terms of $d+1$ massless scalar fields $x^\mu$ freely propagating 
on a two-dimensional Riemannian (~string~) manifold \cite{polya}. By solving 
the field equations for $g_{mn}$ one finds that the intrinsic metric matches
on-shell the induced metric, modulo an arbitrary local rescaling:
\begin{equation}
T^{mn}\equiv \frac{\delta S}{\delta g_{mn}}=0
\longrightarrow g_{mn}(\,\tau ,\sigma\,)\propto \gamma\left(\, x(\,\tau ,\sigma\,)
\,\right)
\label{gind}
\end{equation}
By putting back (\ref{gind}) in (\ref{spol}) one recovers the Nambu-Goto action
again. \\  The action (\ref{spol})
is not only reparametrization invariant but also Weyl invariant, i.e.
one can freely rescale the world manifold metric
\begin{equation}
g_{mn}\longrightarrow \Omega\left(\, \tau\ ,\sigma\,\right)\, g_{mn}
\end{equation}
and, since $x^\mu(\tau,\sigma)$ are fields in a $2D$ spacetime, the freedom due
to Weyl invariance allows to choose a  conformally flat map where:
\begin{equation}
g_{mn}= \eta_{mn}\,\Omega\left(\, \tau^\prime\ , \sigma^\prime\,
\right)\label{confg}
\end{equation}
In this ``conformal gauge'' the string action turns into action of
a $(d+1)$-plet of free scalar fields  in a, two dimensional, flat, spacetime:
\begin{equation}
S=-\frac{\rho}{2}\int_{\mathcal{I}} \ud \tau \: \ud \sigma \: \left(\,
- \dot{x}^2 + x^{\,\prime\, 2}\, \right)
\label{az1}
\end{equation}
where $\dot x^\mu=\partial_\tau x^\mu$, $x^{\prime\mu}=\partial_\sigma x^\mu$.\\
Thus, closed string dynamics in the conformal gauge is described by the
Klein-Gordon field equation supplemented by periodic boundary conditions
\begin{eqnarray}
&&\left[\, -\partial_\tau^2 + \partial_\sigma^2\,\right]\, x^\mu=0\ ,
\label{kg}\\
&& x^\mu\left(\, \tau\ ,\sigma\,\right)=
x^\mu\left(\, \tau\ ,\sigma+\pi\,\right)\label{bc}
\end{eqnarray}
In order to maintain the equivalence with the Nambu-Goto action
one needs to impose the now missing field equations for $g_{mn}$
as \textit{constraints} over the solutions of (\ref{kg}):
\begin{eqnarray}
&& \dot x^\mu x_\mu^\prime =0\label{c1}\\
&& \dot x^\mu \dot x_\mu + x_\mu^\prime x^{\prime\,\mu}=0\label{c2}
\end{eqnarray}

To proceed along the analogy between strings and harmonic oscillators it
is useful to split the string coordinates $x(\tau,\sigma)$ as:
\begin{equation}
x^\mu(\, \tau,\sigma\, )\equiv x_0^\mu(\, \tau\, ) +\eta^\mu (\tau,\sigma)
\label{split}
\end{equation}
The  center of mass coordinates represent the weighted average position of the 
body as a whole. In the string case, one defines $x^\mu _0 (\, \tau\, \,)$ as:
\begin{equation}
x^\mu _0 (\,\tau\,) \equiv\frac{1}{\rho\, \int _0 ^\pi \ud \sigma } \,\rho\,
\int _0 ^\pi \ud \sigma \: x^\mu(\,\tau,\sigma\,) =
\frac{1}{\pi} \int _0 ^\pi \ud \sigma \: x^\mu(\, \tau,\sigma\, )  \ .
\end{equation}
where, the average is taken over the position of each single point along the
string. Thus, the relative coordinates $\eta^\mu (\, \tau,\sigma\,)$ have
vanishing average values 
\begin{equation}
 \int_0^\pi \ud \sigma \: \eta^\mu (\tau,\sigma) =0
\label{etamedio}
\end{equation}
and, in agreement with (\ref{bc}), $\eta^\mu (\tau , \sigma)$ is a periodic 
function in $\sigma$ with
period equal to $\pi$. By inserting the splitting (\ref{split}) in the
action (\ref{az1}) and taking into account (\ref{etamedio}) one gets
\begin{eqnarray}
S&=&S_0 + S_{ho}\label{s0cm}\\
S_0&=&\frac{\pi \rho}{2}\int_0 ^T \ud \tau \:  \dot{x}^2_0\label{scm} \\ 
S_{ho}&=&\frac{\rho}{2}\int_{\mathcal{I}} \ud \tau \: \ud \sigma \:  
\left(\, \dot{\eta}^{\, 2} -\eta^{\,\prime\, 2}\,\right) \label{azcmos}
\end{eqnarray}
The first term (\ref{scm}) represents the  motion of the string center of
mass as the motion of an effective particle of mass $m_0=\pi \rho$. The 
second term (\ref{azcmos}) is reminiscent of harmonic oscillators. To make this
suggestion more explicit, let us Fourier expand the relative coordinate
\begin{equation}
\eta^\mu(\tau,\sigma)=\sum_{n=1}^\infty \left[\, x^\mu_n(\tau) \:
\cos(2n\sigma) + \tilde{x}^\mu_n(\tau) \: \sin(2n\sigma)\,\right]
\end{equation}
where, $x^\mu_n, \tilde{x}^\mu_n \in \mathbf{R}$ for $n=1,\dots,\infty$ 
represent two independent families of
Fourier modes. In terms of the Fourier amplitudes, $\dot\eta^2$ and 
$\eta^{\prime\, 2}$ read:
\begin{eqnarray}
\dot{\eta}^2(\tau,\sigma) &\!\!\!\!\!=&\!\!\!\!\!\!\sum_{n=1}^\infty \sum_{m=1}^\infty 
\left[\,
\dot{x}_n \cdot \dot{x}_m \: \cos(2n\sigma) \cos(2m\sigma) +
\dot{x}_n \cdot \dot{\tilde{x}}_m \: \cos(2n\sigma) 
\sin(2m\sigma) \right. \nonumber \\
& & \!\!\!\left. +\dot{\tilde{x}}_n\cdot \dot{x}_m \: \sin(2n\sigma) 
\cos(2m\sigma)+\dot{\tilde{x}}_n \cdot\dot{\tilde{x}}_m \: 
\sin(2n\sigma) \sin(2m\sigma)\,\right] \\
\eta^{\prime \,2}(\tau,\sigma) &\!\!\!\!\!=& \!\!\!\!\!\! \sum_{n=1}^\infty 
\sum_{m=1}^\infty 4nm
\left[ x_n \;\!\! \cdot x_m  \sin(2n\sigma) \sin(2m\sigma) \! -
x_n \;\!\! \cdot \tilde{x}_m  \sin(2n\sigma) \cos(2m\sigma) \right.\nonumber \\
& & \!\!\!\left. -\tilde{x}_n\cdot x_m \: \cos(2n\sigma) 
\sin(2m\sigma)+\tilde{x}_n\cdot \tilde{x}_m \: 
\cos(2n\sigma) \cos(2m\sigma)\,\right]
\end{eqnarray}
In order to keep notation as simple as possible,  we defined
$ x_n\cdot x_m \equiv \eta_{\nu\mu} x_n^\nu x_m^\mu$. 
Performing integration in $\sigma$ we find
\begin{eqnarray}
\int_0^\pi \ud \sigma \, \dot{\eta}^2&=& \frac{\pi}{2} \sum_{n=1}^\infty
\left(\, \dot{x}^2_n + \dot{\tilde{x}}^2_n\,\right) \\
\int_0^\pi \ud \sigma \, \eta'^2 &=& \frac{\pi}{2} \sum_{n=1}^\infty 4n^2
\left(\, x^2_n + \tilde{x}^2_n\,\right)
\end{eqnarray}
and 
\begin{equation}
S_{ho}= \frac{\rho \pi}{4} \int_0^T \ud \tau \sum_{n=1}^\infty
\left[\, \dot{x}^2_n - 4n^2 x^2_n + \dot{\tilde{x}}^2_n -4n^2 
\tilde{x}^2_n\,\right] \label{lagros}
\end{equation}
Equation (\ref{lagros})  confirms that $S_{ho}$ is the action for
two (~infinite~) families of harmonic oscillators
$x_n(\tau)$ and $\tilde{x}_n(\tau)$ in $d+1$ dimensions.\\
Now, we can conclude that a closed string can be regarded as  a system
of decoupled harmonic oscillators vibrating around a common center of mass
\begin{center}
closed string $=$ point-like center of mass $+$
$x$-harmonic oscillators $ + $
$\tilde{x}$-harmonic oscillators
\end{center}
Indeed, a general solution of (\ref{kg}) with boundary conditions (\ref{bc})
reads
\begin{equation}
x^\mu\left(\, \tau\ ,\sigma\,\right)= x^\mu_0 + \frac{P_0^\mu}{\pi\rho}\tau
+\frac{i}{\sqrt{\pi\,\rho}}\,
\sum_{\genfrac{}{}{0pt}{1}{-\infty}{n\ne 0}}^\infty\frac{1}{2n}
\left[\, \alpha^\mu_n(0) \:
e^{2in(\, \tau-\sigma)} + \tilde\alpha^\mu_n(0) \: e^{2in(\, \tau+\sigma)}\,
\right]
\label{string}
\end{equation}
where, $x^\mu_0$ and $P_0^\mu$ are respectively the initial position and
total momentum of the center of mass. Furthermore, we redefined the oscillator
amplitudes in the following way:
\begin{eqnarray}
&& x_n^\mu(0)\equiv -\frac{1}{n\sqrt{\pi\rho}}Im\left[\, 
\alpha^\mu_n + \tilde\alpha^\mu_n \,\right]\\
&& \tilde{x}_n^\mu(0)\equiv \frac{1}{n\sqrt{\pi\rho}}Re\left[\, \alpha^\mu_n -
\tilde\alpha^\mu_n\,\right]\\
&&\alpha_n^\mu\equiv \alpha^\mu_n(0) \qquad\tilde{\alpha}^\mu_n\equiv
 \tilde\alpha^\mu_n(0)
\end{eqnarray}

In the conformal gauge (\ref{confg}) is easy to show that the string
total momentum is just $P_0$. In fact,
\begin{eqnarray}
  P^\mu &\equiv& \rho\int_0^\pi \ud\sigma \, \dot{x}^\mu\left(\, \tau\
  ,\sigma\,\right)\nonumber\\
  & =&P_0^\mu
\end{eqnarray}
as the contribution from the harmonic oscillators vanishes when integrated
over $\sigma$. Finally, by taking into account the constraints (\ref{c1}),
(\ref{c2}) one obtains the classical form of the energy spectrum
associated to physical, i.e. transverse oscillations (~$k$ runs over $d-1$ 
spatial dimensions~),
\begin{equation} 
E^2= \vec{P}_0^{\, 2} + 4\pi\rho\,\sum_{n=1}^\infty \left(\, \alpha_n^k 
\alpha_{-n\,k} + \tilde{\alpha}_n^k \tilde{\alpha}_{-n\,k}\,\right)
\end{equation} 
 
Upon quantization the classical Fourier amplitudes are replaced by the
occupation quantum numbers and zero-point energy as in (\ref{mass}).
The solution (\ref{string}) and the corresponding
mass spectrum assumes that along any spatial direction
periodic boundary conditions have been imposed. If one, or more, dimension
is compactified, then solution (\ref{string}) has to be modified in a
consistent way and a new symmetry shows up.


\subsection{Toroidal Compactification and $T$-duality at the classical level}

For the sake of simplicity, we are going to consider
the case where only one dimension, say $d^{\mathrm{th}}$, is a 
``\textit{circle}'' of radius $R$.
Thus, the position along $x^d$ is defined modulo $2\pi\, R$: $x^d\simeq
x^d + 2\pi R $. $x^d$ must solve the string wave equation by taking
into account such a new boundary condition
\begin{eqnarray}
&&\ddot x^d- x^d{}^{\, \prime\prime}=0\ ,\\
&& x^d\left(\, \tau\ , \sigma +\pi\,\right)=
x^d\left(\, \tau\ , \sigma\,\right) + 2\pi\, w\, R\ , \qquad w=0\ ,\pm 1\ ,
\pm 2\ ,\dots
\end{eqnarray}
The integer $w$ is the \textit{winding number}, i.e. the number of times the
string wraps around $x^5$.
The solution (\ref{string}) becomes now
\begin{eqnarray}
x^d\left(\tau , \sigma\right)
&\!\!=&\!\! x^{d}_0 +\frac{n}{\pi\, \rho\, R}\, \tau + 2 w\, R\,\sigma\nonumber\\
   &\!\!+&\!\!  \frac{i}{2\sqrt{\pi \rho}}\sum_{m\ne 0} \frac{1}{m}
    \left[ \alpha_m^{d}\exp\left(
   -2im\left( \tau -\sigma \right) \right)  +  
   \widetilde\alpha_m^{d}\exp\left(
   -2im\left( \tau +\sigma \right) \right) \right] \nonumber\\
&& 
\end{eqnarray}
where we have taken into account the quantization relation (\ref{db}).
Furthermore, from equation (\ref{az1}), we find:
\begin{eqnarray}
\mathrm{P}^{d}&\!\equiv&\! \rho\, \dot x^d\nonumber\\
&\!=&\! \frac{n}{\pi \, R} +\sqrt{\frac{ \rho}{\pi}}\sum_{m\ne 0}\,
   \left[\, \alpha_m^{d}\exp\left(\,
   -2im\left(\, \tau -\sigma\,\right)\,\right)  +  
   \widetilde\alpha_m^{d}\exp\left(\,
   -2im\left(\, \tau +\sigma\,\right)\,\right) \,\right]\nonumber\\
   &&
\end{eqnarray}
We recall that $n$ is the Kaluza-Klein quantum number.
The resulting forms of energy and mass spectra are respectively
\begin{eqnarray}
 E^2_{n\, w\, N\, \tilde{N}}&\!\!=\!\!& P_0^k P_{0\, k}+
\frac{n^2}{ R^2} +4\pi^2 \rho^2\, w^2\, R^2
+2\pi \rho \left( N +\tilde{N} -\frac{d-1}{12} \right) \label{enwNN}\\
M^2_{n\, w\, N\, \tilde{N}} &\!\!=\!\!& 4\pi^2 \rho^2\, w^2\, R^2
+2\pi \rho\,\left(\, N +\tilde{N} -\frac{d-1}{12}\,\right)
\end{eqnarray}
and the index $k=1\ ,2\ ,\dots d-1$ labels non-compact dimensions.
In the expression (\ref{enwNN}) above the first term is the contribution from the 
spatial continuous momentum; the second term comes from the discretized momentum 
along the fifth dimension; the third term is the winding which accounts for 
the wrapping of the string around $x^d$, and the last term describes the 
oscillators with their spectra and zero energy.\\ 
The four quantum numbers $N$, $\tilde{N}$, $n$, $w$ are not independent.
Indeed string constraints require a level matching condition \cite{book}
\begin{equation}
\tilde{N}- N = nw
\end{equation}
Thus, both energy and mass spectrum can be re-written as
\begin{eqnarray}
E^2_{n\, w\, N}
&=& P_0^k P_{0\, k}+\frac{n^2}{R^2} +\frac{w^2\, R^2}{\alpha^{\prime\,2}}+
\frac{1}{\alpha^\prime}\,\left(\,2N +nw  -\frac{d-1}{12}\,\right)
\label{spectre1}\\
M^2_{n\, w\, N}
&=& \frac{w^2\, R^2}{\alpha^{\prime\, 2}}+
\frac{1}{\alpha^\prime}\,\left(\,2N +nw  -\frac{d-1}{12}\,\right)
\label{spectre2}
\end{eqnarray}
Now one can read that
the total energy of the system is the sum of contributions from
winding modes, Kaluza-Klein modes, oscillators and zero-point energy
and it is manifest that 
string energy spectrum is \textit{invariant under $T$-duality 
transformation}, as we discussed in the previous section.


\section{Effective String propagator}

In the previous section we introduced strings as 
infinite collections of relativistic harmonic oscillators. 
Thus, one could
start quantization from the equivalent action (\ref{lagros}). But,
in order to quantize physical degrees of freedom only, we need to remove
fictitious harmonic components required to keep the formulation manifestly
Lorentz invariant. Accordingly, let us take a step backward and consider
the split Lagrangian
\begin{equation} 
L= \frac{\pi\rho}{2}\dot x^\mu_0 \dot x_{0\, \mu} +\frac{\rho}{2} 
\int_0^\pi \ud\sigma
\left[\,\dot{\eta}_\mu\, \dot{\eta}^\mu -{\eta}^\prime_\mu\, 
{\eta}^{\prime\, \mu}\,\right]\label{lspit}
\end{equation} 
The Lagrangian (\ref{lspit}) is a Lorentz scalar hence we are free
to choose the most appropriate inertial frame. A specially useful choice
is the \textit{light-cone frame} where the timelike and one of the space-like
axis are rotated along the light-cone:
\begin{equation}
x^\pm =\frac{x^0\pm x^1}{\sqrt{2}} 
\end{equation}
In such a coordinate frame Lagrangian (\ref{lspit}) reads
\begin{equation} 
L= \frac{\pi\rho}{2}
\left(\,-2 \dot{x}^+_0 \dot{x}^- _0 + \dot x^i_0 \dot x_{0\, i}\,\right) 
+\frac{\rho}{2} \int_0^\pi \ud\sigma
\left[\,  -2\dot{\eta}^+ \dot{\eta}^-  +\dot{\eta}_i\, \dot{\eta}^i  
+2\eta^{\prime\, +} \eta^{\prime\, -} 
 -{\eta}^\prime_i\, {\eta}^{\prime\, i}\,\right]\label{llc}
\end{equation} 
where the index $i$ labels $d-1$ ``transverse'' space-like directions.\\
Now, having  the action we can calculate the kernel in
the path integral formulation. We can formally write down the form
of the transition
amplitude from an initial to a final configuration of the whole system as a path
integral
\begin{equation}
\langle f\vert i\rangle \equiv \int_{x_{0,i}}^{x_{0,f}} [Dx_0]\, 
\int_{\eta_i}^{\eta_f} 
[D\eta] \: \exp \left[ i \int_0^T \ud\tau\, 
L\left(\, \dot{x}_0\ , x_0\ ; \dot\eta \ , \eta \,\right) \right]
\end{equation}
where $x_{0,i}$ and $x_{0,f}$ represent initial and final position of the
string center of mass, while $\eta_i\equiv \eta(\, 0,\sigma)$ and
$\eta_f\equiv \eta(\, T,\sigma)$ are initial and final configuration
of the fluctuating part of the string.\\
We are not really looking for this quantity, rather we look for the
analogue expression of the Kaluza-Klein effective propagator (\ref{kernel})
once the pointlike probe is replaced by a string. In other words, we
are going to recover the effective propagator for the string center of
mass propagating in vacuum where all possible string fluctuations take place.
To account for virtual transitions among string states means to sum over
closed paths in the $\eta$ configurations space, i.e. the $\eta$ to be summed
over satisfy the boundary condition $\eta_i=\eta_f$. This leads to the
 following expression
\begin{equation}
Z\left(\, T\,\right) \equiv \oint [D\eta] 
\: \exp \left[ i \int_0^T \ud\tau\, L \left(\dot\eta \ , \eta \right) \right]
\label{zeta}
\end{equation}
where $L(\dot{\eta},\eta)$ represents the second term on the right hand
side of equation (\ref{llc}).
This kind of path integral over closed paths is often called ``vacuum
partition functional'' to remark the formal analogy between (\ref{zeta}) and the 
statistical mechanics partition function. In this respect it is important to 
notice that $Z$ is not describing a physical gas of strings \cite{gas},
i.e, a fluid of real objects.
Rather it can be seen as a mathematical quantity encoding the property
of the string physical vacuum where all kind of virtual transitions take place.
In order to compute  $Z$ in the correct way,
we have to remove unphysical modes. This can be done by choosing  the
\textit{light-cone gauge}, which is a frame where
all the oscillations along $+$ direction are turned-off
\begin{eqnarray}
\eta^+ &=&0\\
\dot{\eta}^+ &=&0\\
\eta^{\prime\, +}&=&0
\end{eqnarray}
Thus, in the light-cone  gauge the partition functional reads
\begin{equation}
Z\left(\, T\,\right)=\oint [\,D\vec\eta\, ] \: \exp \left[\, i\frac{\rho}{2} 
\int_0^T \int_0^\pi \ud\tau\, \ud\sigma \left(\,\dot{\vec{\eta}}^{\, 2} -
 \vec{\eta}^{\, \prime \, 2}\,\right)\, \right]
\label{zetalc}
\end{equation}
where only transverse physical oscillations are summed over. 
In analogy to the Coulomb gauge in electrodynamics, the light-cone
gauge allows to remove both
``timelike'' $\eta^+$ and ``longitudinal'' $\eta^-$ components of the
$\eta$-field.\\
  At this point we use the Fourier expansion for the transverse
coordinates to write $Z\left(\, T\,\right)$ in the form of a partition functional
 for two infinite families of transverse harmonic oscillators 
\begin{eqnarray}
Z\left(\, T\,\right)&=&\oint \prod _{n=1} ^\infty \ud
\vec{x}_n
\prod _{n=1} ^\infty \ud \tilde{\vec{x}}_n \: \exp \bigg[i \frac{\rho \pi}{4} 
\int_0^T \ud \tau \sum_{n=1}^\infty
\big[ (\dot{\vec{x}}^{\, 2}_n - 4n^2 \vec{x}^{\, 2}_n) \nonumber\\ 
&\qquad \qquad \vec{\phantom{x}} \!\!\!\!\!\!\!\!& \!\!\!\!\!\!\!\!\!\!\! \: 
+  ( \dot{ \tilde{\vec{x}}  }^{\, 2}_n -4n^2 \tilde{\vec{x}}^{\, 2}_n ) 
\big] \bigg] 
\label{Kos}
\end{eqnarray}
Thus, the string partition function turns out to be an infinite product
of harmonic oscillator  path integrals computed over families of closed paths.
For an elementary harmonic oscillator the partition functional is known and
reads
\begin{equation}
Z_{ho}\left(\, t\, \right)=\sum_{n=0}^\infty \exp \left[\,- i\, t \,
\omega_0\,\left(\, n +\frac{1}{2}\,\right)\,\right]
\end{equation}
where, $\omega_0$ is the fundamental frequency and $n$ is the principal
quantum number. By considering the string as an infinite collection of 
harmonic oscillators, we find 
\begin{equation}
Z_{ho}\left(\, T\,\right)=\sum_{N,\tilde{N}=0}^\infty \exp 
\left[\,- i \, T \,
\left(\,N +\tilde{N}-\frac{d-1}{12}\,\right)\,\right]
\end{equation}
 
Whenever one (~ or more~) dimension is compact,  strings
can wrap around it an arbitrary number of times. Accordingly, we have to 
correct
the string spectrum by taking into account the contribution from the winding
modes. For the sake of simplicity, let us consider again the case of a single
compact dimension. Then, we find
\begin{equation}
Z_{ho}\left(\, T\ , R\,\right)=
\sum_{N,\tilde{N},w =0}^\infty \exp \left[\, -i\, T \,\left(\, N +
\tilde{N}-\frac{d-1}{12} + w^2\,\frac{R^2}{\alpha ^\prime }\,\right)\,
\right]\label{zw}
\end{equation}
By including winding modes in (\ref{zw}) we encode a topological feature
which makes the string substantially different from a pure ``gas'' of 
pointlike oscillators.\\
Now we can give a more definite meaning to the  center of mass kernel
in the vacuum which is filled up with both Kaluza-Klein type fluctuations
(~we described in Section(2)~) and the new kind of virtual processes brought
in by the string excitation modes. By putting all together we find 
\begin{eqnarray}
K\left(\, x_{0, f} -x_{0,i}\ ; T\,\right)&=&\sum_{N,\tilde{N},w =0}^\infty \exp
\left[\, -i T \,\left(\,N +
\tilde{N}-\frac{d-1}{12} + w^2\,\frac{R^2}{\alpha^\prime} \,\right)\,
\right] \times \nonumber \\
&&\int_{ x_0(0)= x_{0,i}  }^{ x_0(T)=x_ {0, f} } \left[\, Dx_0\,\right] 
\: \exp \left[ i \int_0^T \ud\tau\, 
L\left(\, \dot{x}_0\ , x_0 \,\right) \right]
\label{kcm}
\end{eqnarray}
 
Let us consider the center of mass path integral in more detail. 
In analogy to the oscillator part calculation we pass to the light cone gauge,
writing explicitly:
\begin{equation}
x_0^\mu(\,\tau\,)=\left(\, x_0^+(\,\tau\,)\ , x_0^-(\,\tau\,)\ ,
 x_0^j(\,\tau\,)\ , x_0^d(\,\tau\,) \, \right)
\end{equation}
where, $j=2\ , 3\ ,\dots d-1$ runs over non-compact dimensions and we have
chosen the last one, namely $x^d$, to be compact. It follows 
that integration measure reads:
\begin{equation}
\int_{x_0(0)=x_{0, i}}^{ x_0(T)=x_{0, f}} \left[\, Dx_0\, \right]=
\int_{x_{0, i}^+}^{x_{0, f}^+} \left[\, Dx_0^+\, \right]
\int_{x_{0,i}^-}^{x_{0, f}^-} \left[\, Dx_0^-\, \right] 
\int_{\vec{x}_{0, i}}^{\vec x_{0,f}} \left[\, D\vec x_0\, \right] 
\int_{x_{0, i}^d}^{x_{0, i}^d + n\,l_0} \!\!\!\!\!\! \left[\, Dx_0^d\, \right] 
\end{equation}
From now on, we shall omit the subscript $0$ to indicate that we are
referring to center of mass because no ambiguity with oscillators coordinates
and momenta can occur.\\
The technically simpler way to compute (\ref{kcm}) is by weighting each 
path by its canonical action in phase-space as done in Section (2):
\begin{equation}
S=\int_0^T \! \ud\tau \left[ P_+ \dot{x}^+ \!+ P_- \dot{x}^- \!
+ P_j \dot{x}^j + P_d \dot{x}^d - \frac{1}{2 \pi \rho} \left( 2 P_+P_-  
+ P_j P^j +P_d^2 \right) \right]
\label{scan}
\end{equation}
%
Repeating the calculations of Section(2), and dropping both $n=0$ and
$w=0$ modes,  we obtain
\begin{eqnarray}
K_{reg}\left(\, x_f-x_i\ ; T\,\right)&=&  
\left( \frac{1}{4 i\pi \alpha^\prime T}\right)^\frac{d-1}{2} 
\sum_{\genfrac{}{}{0pt}{1}{N=0}{w,n=1}}^\infty \exp \left[
-\frac{ \left( x_f - x_i \right)^2 + n^2 l_0^2]}{4i
\alpha^\prime T} \right] \times \nonumber\\
&& \exp \left[ - i T\left(2N + nw
 -\frac{d-1}{12} + \frac{w^2 R^2}{\alpha^\prime}\right)\right]
\label{Ksreg}
\end{eqnarray}
From (\ref{Ksreg}) it is possible to obtain the Green function by
integration over the unmeasurable lapse of time $T$ as follows
\begin{eqnarray}
G\left( x_f-x_i \right) &\equiv& \left( 2\pi \right)^\frac{d-1}{2}
\int_0^\infty \ud T e^{-i2\alpha^\prime m_0^2 T} 
K_{reg} \left( x_f - x_i ; \right) \\
&=&\left(\frac{1}{2i\alpha^\prime}\right)^\frac{d-1}{2} 
\sum_{\genfrac{}{}{0pt}{1}{N=0}{w,n=1}}^\infty \int_0^\infty \ud T \, 
T^{-\frac{d-1}{2}} \exp\left[ -\frac{ \left( x_f - x_i \right)^2 + n^2 l_0^2}
{4i\alpha^\prime T} \right] \times \nonumber\\
&& \exp \left[ -i T\left(2\alpha^\prime m_0^2 + 2N + nw
 -\frac{d-1}{12} + \frac{w^2 R^2}{\alpha^\prime}\right)\right]
\label{greall}
\end{eqnarray}
where $m_0$ can be zero or non vanishing and has been introduced to account for 
low energy effects, e.g. spontaneous symmetry breaking.\\
In order to evaluate the short distance behavior of the Green function
(\ref{greall}) it is useful to Fourier transform it
\begin{eqnarray}
G\left( p \right) &\equiv& \int \ud^{d-1} u \left( 2 \pi \right)
^{-\frac{d-1}{2}} \, e^{i p u} \, G\left( u \right) \nonumber \\
&=& \sum_{\genfrac{}{}{0pt}{1}{N=0}{w,n=1}}^\infty \int_0^\infty \ud T \, 
\exp\left[- i \left( p^2 + M_{N,w,n}^2 \right) T - \frac{n^2 l_0^2}{4i
T} \right] \\
&=&\sum_{\genfrac{}{}{0pt}{1}{N=0}{w,n=1}}^\infty 
\frac{n l_0}{\sqrt{p^2 + M_{N,w,n}^2}}
\, K_1 \left( n l_0 \sqrt{p^2 + M_{N,w,n}^2}\right)
\label{grep}
\end{eqnarray}
where in the second step we rescaled $T \to \alpha^\prime T$ while, in the third, we rotated 
to Euclidean space. The mass term $M_{N,w,n}^2$ is defined as 
\begin{equation}
M_{N,w,n}^2 \equiv \frac{1}{\alpha^\prime} \left(  
 2N + nw -\frac{d-1}{2} + \frac{w^2 R^2}{\alpha^\prime} +
  2 \alpha^\prime m_0^2 \right)
\label{MNwn}
\end{equation}
The result we just found is, as expected, the same as in (\ref{gpreg}) with the
replacement $m_0^2 \to M_{N,w,n}^2$, and the choice of the leading term $N=0$, $w=1$, $n=1$ 
(~$d=4$ and there's only one compact extra-dimension~).\\
A generalization to more than one compact dimensions is straightforward: one has
only to perform the replacements
\begin{equation}
n \to n_i \quad \left(\Rightarrow n^2 \to \sum_i n_i^2\right)\, , \qquad 
w \to w_i \quad \left(\Rightarrow w^2 \to \sum_i w_i^2\right) 
\end{equation} 
and obviously
\begin{equation}
R \to R_i \quad \left(\Rightarrow R^2 \to \sum_i R_i^2\right)\, , \qquad 
l_0 \to l_{0,i} \quad \left(\Rightarrow l_0^2 \to \sum_i l_{0,i}^2\right) 
\end{equation}
where the index $i$ counts the compact extra-dimensions.


\section{Conclusions and Outlook}

The need of more than four dimensions in order to build up a consistent 
unified
theory of fundamental interactions is now generally accepted, irrespectively of
the particular unification framework one chooses.\\
The problem of reducing the higher dimensional world to our physical spacetime
is solved by introducing the concept of compact dimensions. If the length 
scale
characterizing the compact dimensions is comparable with the Planck length,
then, there is no way to detect the extra-dimensions at the energy presently
available in particle accelerators. Kaluza-Klein, or string, excitation modes
require $E\approx \sqrt{\alpha^\prime}\approx 10^{19}GeV$ for being produced 
in high-energy scattering events. This reasoning understands to produce 
Kaluza-Klein particles, or excited strings, as \textit{real} objects capable to 
leave a track,
or produce a signal, in some appropriate revelation device.\\
In this paper we faced the problem from a different perspective: in 
alternative to the production of real objects sensitive to the presence of 
extra-dimensions,
we focused on the \textit{virtual} presence of these objects in the four 
dimensional vacuum. Indeed, the Uncertainty Principle allows the ``spooky''
presence of any state in the quantum vacuum, irrespectively of the energy
needed to produce them as real, detectable, objects.\\
We have discussed a model of four dimensional vacuum where stringy 
fluctuations
take place in form of Kaluza-Klein, winding and harmonic oscillators
virtual excitations. The total effect amounts to introduce a length scale
$l_0 = 2\pi\sqrt{\alpha^\prime}$ as a fundamental property of this new
kind of vacuum. The presence  of $l_0$ is remarked by the asymptotic
behavior of Feynman propagators, where $l_0$ acts as an exponential 
ultra-violet cut-off. 
The conceptual crux leading to the suppression of ultraviolet infinities is
the choice of a particular kind of paths in the functional integral: we 
selected
histories which are closed along the extra-dimensions. Thus, the total path 
integral factorizes into the product of the four dimensional path integral
for the propagator times the vacuum partition functional accounting for 
virtual
fluctuations along extra-dimensions. On the technical side, self-consistency
requires dropping the zero-modes, which describe topologically trivial 
fluctuations. We mean, fluctuations described by paths which can be 
continuously
shrunk to a point. With hindsight, it is clear how zero-modes bring 
ultraviolet divergences in, as they are ``blind'' to the extra-dimensions and
can probe arbitrary short distance.\\
Against this background, we propose a model for the physical vacuum where
only topologically non-trivial fluctuations take place. Anyway, the standard 
Minkowski vacuum, with its pathological short-distance behavior, can be 
recovered again, in the limit $l_0\to 0$. In other words, standard quantum
field theory ultraviolet infinities show up again as soon as we remove the 
cut-off $l_0$, as it is expected. 
It can be worth observing that in our formulation the limit $l_0\to 0$ is 
equivalent to the infinite tension limit, i.e. $\alpha^\prime\to 0 $, where 
strings shrink to structureless points and the point-particle picture
of matter is recovered.\\
We can summarize the main results of our approach as follows. Zero-point 
length in four dimensional spacetime can be seen as the \textit{virtual memory} of
the presence of compact extra-dimensions even much below the threshold energy 
needed to produce real Kaluza-Klein particles. 
If one sticks to a Kaluza-Klein quantum field 
theory picture the actual value of $l_0$ remains undetermined. In the more
general framework provided by string theory, $T$-duality selects the 
\textit{unique} self-dual value for the compactification scale, and, 
accordingly determines $l_0=2\pi\sqrt{\alpha^\prime}$.\\
We are aware that our approach does not fit into the ``current wisdom'' where
zero-modes are \textit{tour court} assumed to describe the four dimensional world.
Our point of view is slightly different. Following the tenets of Quantum Mechanics
we considered the effects of virtual sub-threshold string excitations in the
physical vacuum, and recovered the presence of a minimal length as a memory of
the higher dimensional spacetime where string theory is defined.\\
If our model is correct, then, one could expect to see deviations from the
theoretical predictions of standard quantum field theory due to the presence
of the modified Feynman propagator (\ref{grep}). Those deviations should appear
in an intermediate energy regime much below the string scale and, hopefully,
not too much over the energy range presently available. In this respect, 
the most optimistic scenario is offered by the ``$TeV$ scale unification'' 
models, where the string scale is lowered down to some $TeV$ 
\cite{tev},\cite{koko}.   
Such a ``low-energy'' unification can be realized provided the 
extra-dimensions are compactified to a ``large'' radius of some fraction of 
millimeter. In this case $l_0\approx 10^{-17}cm.$ and its presence would be 
detectable in the high-energy scattering experiments \cite{sabine} planned for 
the next generation of particle accelerators.


\section{Acknowledgments}

One of the authors, (E.S.), would like to thank Prof. T. Padmanabhan
for ``triggering'' his interest to find out a link between zero-point
length and string theory \cite{unlucky};  the same author would also like to
thank Dr. S. Shankaranarayanan for useful discussions and
pointing out Ref.\cite{balbi}.


\begin{thebibliography}{99}
\bibitem{yoneya} T. Yoneya  Prog. Theor. Phys. \textbf{56}, 1310 (1976)
\bibitem{snyder} H. S. Snyder  Phys. Rev. \textbf{71}, 38 (1947)
\bibitem{ncqft} A. Smailagic, E. Spallucci  J. Phys. A \textbf{37},   7169 
(2004)
\bibitem{garay} L.J. Garay  Int. J. Mod. Phys. \textbf{A10}, 145 (1995) 
\bibitem{padma1} T. Padmanabhan  Ann. Phys. (N.Y.) \textbf{165}, 38 (1985);\\
T. Padmanabhan  Class. Quantum Grav. \textbf{4}, L107 (1987)
\bibitem{balbi} R. Balbinot, A. Barletta  Phys. Rev. D \textbf{34}, 3666 (1986)
\bibitem{fradefi} E.S. Fradkin Nucl. Phys. \textbf{49}, 624 (1963);\\ 
G.V. Efimov Sov. Phys. JETP \textbf{17}, 1417 (1963);\\
A. Salam  Coral Cable conference (London 1970)
\bibitem{paddy} J.V. Narlikar, T. Padmanabhan
\textit{Gravity, Gauge Theories and Quantum Cosmology}, D. Reidel Publishing
Company, 1986
\bibitem{del} R. Delbourgo  Rep. Progr. in Phys. \textbf{39}, 345 (1976)
\bibitem{salam} C.J. Isham, A. Salam, J. Strathdee  Phys. Rev.\textbf{3} (1971)
\bibitem{padma2} T. Padmanabhan  Phys. Rev. Lett. \textbf{78},  1854 (1997);\\
T. Padmanabhan  Phys. Rev. D. \textbf{57}, 6206 (1998). 
\bibitem{shanky} S. Shankaranarayanan, T. Padmanabhan  Int. J. Mod. Phys. 
\textbf{D10}, 351 (2001)
\bibitem{rebbimand} C. Rebbi  Phys. Rept. \textbf{12}, 1 (1974)\\
S. Mandelstam  Phys. Rept. \textbf{13}, 259 (1974)
\bibitem{greschwa} M.B. Green, J.H. Schwarz  Phys. Lett. \textbf{B109}, 444 
(1982)
\bibitem{kaluza} T. Kaluza  Sitzungsber. Preuss. Akad. Wiss. Berlin (Math. 
Phys.) \textbf{K1}, 966 (1921)
\bibitem{klein} O. Klein  Z. Phys. \textbf{37}, 895 (1926);\\
O. Klein  Surv. HEP \textbf{5}, 241 (1986)
\bibitem{collins} P.D.B. Collins  \textit{Introduction to Regge Theory and 
High Energy Physics}, Cambridge University Press, 1977;\\
P.D.B. Collins, E.J. Squires  \textit{Regge Poles in Particle Physics},
Springer-Verlag, 1968
\bibitem{feyn} P. Feynman, A.R. Hibbs \textit{Quantum Mechanics and Path
Integrals}, McGraw-Hill Companies, 1965
\bibitem{kleinert} H. Kleinert  \textit{Gauge Fields in Condensed Matter}, 
World Scientific Pub. Co. Inc., 1990
\bibitem{ej} S. Ansoldi, A. Aurilia, E. Spallucci  Eur.J.Phys. \textbf{21}, 
1 (2000) 
\bibitem{nambu} T.Goto, Progr. Theor. Phys. \textbf{46} 1560 (1971);\\
Y.Nambu   Phys. Rev. \textbf{D}10: 4262 (1974)
 \bibitem{polya} A.M. Polyakov, Phys. Lett.\textbf{B}103, 207  (1981);
 ibidem,  211,  (1981)
\bibitem{candelas} P. Candelas, S. Weinberg Nucl. Phys. B \textbf{237}, 397 
(1984)
\bibitem{schwi} J.S. Schwinger  \textit{Particles, Sources, and Fields},
Addison-Wesley Pub. Co., 1970
\bibitem{book} 
B.\ Hatfield, \ \textit{Quantum field theory of point particles and strings}\ 
Addison-Wesley,\  (1992)\\
J.\  Polchinski \textit{String Theory, Vol. 1}
Cambridge Monographs on Mathematical Physics
\bibitem{gas} S. Jain \textit{Absence of initial singularities in superstring cosmology}
 \hbox{gr-qc/9708018;} \\
S. P. Patil, Robert Brandenberger \textit{Radion Stabilization by Stringy Effects 
in General Relativity and Dilaton Gravity}, hep-th/0401037
\bibitem{tev}  I. Antoniadis, N. Arkani-Hamed, S. Dimopoulos, G. Dvali
Phys.\ Lett.\ \textbf{B436}\  (1998)\  257\\
G. Shiu, S.-H. Henry Tye  Phys.\ Rev.\ \textbf{D58}\  (1998)\  106007\\
N. Arkani-Hamed, S. Dimopoulos, G. Dvali
Phys.\ Rev.\  \textbf{D59}\ (1999)\ 086004
\bibitem{koko} D. Cremades, L. E. Ibanez, F. Marchesano\
Nucl.\ Phys.\ \textbf{B643},  93  (2002); \\
C. Kokorelis\ Nucl.\ Phys. \ \textbf{B677},  115 (2004)
\bibitem{sabine} U. Harbach, S. Hossenfelder, M. Bleicher, H.
Stoecker,\ Phys. Lett. \textbf{B584}, 109 (2004)\\
S. Hossenfelder,\   Phys.\ Lett.\ \textbf {B598},  92 (2004)
\\
S. Hossenfelder,\ Phys.\ Rev.\  \textbf{D70},  105003 (2004)
\bibitem{unlucky} A. Smailagic, E. Spallucci, T. Padmanabhan
\textit{String theory $T$-duality and the zero point length of spacetime},
hep-th/0308122 
\end{thebibliography}
\end{document}